\documentclass[letterpaper]{article}
\usepackage{aaai}
\usepackage{times}
\usepackage{helvet}
\usepackage{courier}
\usepackage{graphicx}
\usepackage{amsmath}
\usepackage{amsthm}
\usepackage{mathrsfs}
\usepackage{color}
\usepackage{listings}
\usepackage{subfigure}
\newtheorem{myTheorem}{Theorem}
\newtheorem{myProof}{Proof}

\newtheorem{myDefinition}{Definition}

\newtheorem{myLemma}{Lemma}

\newtheorem{myCorollary}{Corollary}
\begin{document}
%
\title{Invincible Strategies of Iterated Prisoner's Dilemma}
\author{Shiheng Wang \and Fangzhen Lin\\
Department of Computer Science\\
The Hong Kong University of Science and Technology\\
Clear Water Bay,Kowloon,Hong Kong}
\maketitle

\begin{abstract}
\begin{quote}
Iterated Prisoner's Dilemma (IPD) is a well-known benchmark for studying the long term behaviours of rational agents, such as how cooperation can emerge among selfish and unrelated agents that need to co-exist over long term. Many well-known strategies have been studied, form the simple tit-for-tat (TFT) strategy made famous by Axelrod after his influential tournaments to more involved ones like zero determinant strategies studied recently by Press and Dyson. In this paper, following Press and Dyson, we consider one memory probabilistic strategies. We consider what we call invincible strategies: a strategy is invincible if it never loses against any other strategy in terms of average payoff in the limit, if the limit exists. We show a strategy is invincible iff $p_4=0$ and $p_2+p_3\leq 1$, where $p_4$ is the probability of playing $C$ given that the profile
in the previous round is $(D,D)$ (meaning both defected in the last round), and similarly $p_2$ and $p_3$ are the probability of playing $C$ when the profiles in the previous round are $(C,D)$ and $(D,C)$, respectively.
\end{quote}
\end{abstract}

\section{Introduction}
    Iterated Prisoner's Dilemma is a classical problem to understand rational agents' behavior. It involves two agents playing repeatedly the Prisoner's Dilemma (PD). In PD, each player can choose between Cooperate (C) and Defect (D). If both choose C, they receive a payoff of R (rewards); If both choose D, they receive a payoff of P (penalty); If one chooses C and the other D, the defector receives a payoff of T (temptation to defect) and the cooperator a payoff of S (socker's payoff).     Table \ref{PD} gives a normal form representation of this game \cite{rapoport1965prisoner}.
    \begin{table}[htbp]
      \centering
        \begin{tabular}{|c|c|c|}
          \hline
            &   C   & D \\
          \hline
          C & (C,C) & (S,T) \\
          \hline
          D & (T,S) & (P,P) \\
          \hline
        \end{tabular}
      \caption{Prisoner's Dilemma}\label{PD}
    \end{table}

Under the assumption that $T>R>P>S$, the profile $(D,D)$ is
the dominant Nash equilibrium of the game. However, both would receive a higher
payoff of $R$ if they decide to cooperate.
There is no controversy what a rational agent would do when playing PD. However, if the game is repeated indefintely, it is not clear which if any strategy is the best. In fact, researchers from diverse disciplines have used the IPD to study the emergence of cooperation among unrelated agents.
Robert Axelrod was the first to run computer tournaments on iterated prisoner's dilemma. Surprisingly, a simple Tit-For-Tat strategy won the tournaments.
Axelrod concluded that this simple cooperative strategy won them for good reasons, and, together with his collaborators, applied this observation to evolutionary
biology, among other things \cite{hamilton1981evolution}.
In 2012 Press and Dyson \cite{press2012iterated} dramatically changed people's understanding of this problem by deriving what they called zero determinant (ZD) strategies. Of particular interests is a subclass of these strategies called extortionate
strategies that can
enforce an extortionate linear relation between the players' scores.
After the discovery of ZD strategies, Stewart and Plotkin \cite{stewart2012extortion} did another computer tournament with the addition of some ZD strategies.
The results were that ZD strategies are similar to mean strategies like
\emph{Always Defect}. While they would won most number of games but still
got low average payoffs. Generous strategies like TFT still have higher
average payoffs. Hilbe {\it et al.} \cite{hilbe2013evolution} did some
experiments and concluded that with large enough populations, extortionate strategies
can actually act as catalysts for the evolution of cooperation, similar to tit-for-tat, but that they are not the stable outcome of natural selection.

    An agent is in dilemma when faced with an extortionate co-player. If it chooses to cooperate, it automatically accepts the extortion and gives the co-player more payoff. If it defects, itself will receive a punishment payoff of P. While studying whether there is any way to retaliate against the extortioners, we disappointedly find that extortioners are actually invincible. This invincible property guarantees the success of extortion, thus foes cannot think out any strategy to revenge.

    In this paper we study a class of strategies called invincible strategies. Such strategies would never lose against any other strategies in terms of
limit of average payoffs. We'll see that these strategies include many well-known
ones such as TFT, Always Defect and extortionate strategies.
The characterization of these strategies turned out to be surprisingly
concise, and
we derive them by modeling iterated prisoner's dilemma as a Markov process as in \cite{press2012iterated}.

    The rest of this paper is organized as follows. We first introduce iterated prisoner's dilemma and important tournaments. Next we make a brief review of Press and Dyson's work of zero determinant strategies. Following that we give the characteristic and proof of invincible strategy, then we describe some experimental results on these strategies.

\section{Background}
\subsection{Iterated Prisoner's Dilemma}

As mentioned, IPD is just repeated PD given in Table~\ref{PD}, with the
following restrictions on the payoff numbers:
    \begin{itemize}
      \item $T>R>P>S$. This makes (D,D) the dominant equilibrium in PD.
      \item $2R>T+S$. This makes (C,C) the best payoff for both players in the long run.
    \end{itemize}
    One concrete example of Prisoner's Dilemma is,
    \begin{table}[htbp]
      \centering
        \begin{tabular}{|c|c|c|}
          \hline
            &   C   & D \\
          \hline
          C & (3,3) & (0,5) \\
          \hline
          D & (5,0) & (1,1) \\
          \hline
        \end{tabular}
      \caption{Prison's Dilemma Game (concrete example)}\label{Specific_PD}
    \end{table}

    The Prisoner’s Dilemma itself is well established as a way to study the emergence of cooperative behavior. If the Prisoner’s Dilemma is played only once, it always pays to defect - even though both players would benefit by both cooperating. If the game is played more than once, however, other strategies, that reward cooperation and punish defection, can dominate the defectors, especially when played in a spatial context or for an indeterminate number of rounds\cite{alexrod1983evolution}.

 We  calculate the payoff of iterated prisoner's dilemma by average payoff
in the limit (cf. \cite{shoham2008multiagent}):
      Given an infinite sequence of payoffs $r_i^{(1)},r_i^{(2)},...$ for player i, the average payoff of i is
      $$  \bar{r} = \lim_{k \to \infty } \frac{\sum_{j=1}^{k}r_i^{(j)}}{k}  $$

\subsection{Tournaments of IPD}
As mentioned, in Axelrod's landmark tournaments of iterated prisoner's dilemma \cite{alexrod1983evolution},
 Tit-For-Tat proved to be an extraordinarily successful way to foster cooperation and accumulate a large payoff. Tit-For-Tat is a strategy which cooperates if his opponent cooperates on the last round, and otherwise defect. This tournament inspires much discussion about Iterated Prisoner's Dilemma (see, e.g. \cite{kendall2007iterated}).

    In order to test whether zero determinant strategies enjoy advantage in Axelrod's original tournament, Stewart and Plotkin\cite{stewart2012extortion} reran that competition with the addition of some ZD strategies in 2012. Extortionate Strategy Extort-2 forces the relationship $s_X-P = 2(s_Y-P)$, where $s_X$ and $s_Y$ represent the players' scores. On the other hand, Generous Strategy ZDGTFT-2 forces the relationship $s_X-R = 2(s_Y-R)$ between the players' scores. ZDGTFT-2 is a more generous strategy, which offers its opponent a higher portion of payoffs above P.

    Two metrics are considered for this tournament, i.e. average payoff and wins of pairwise matches. In terms of average payoff, ZDGTFT-2 receives the highest score, higher even than Tit-For-Tat and Generous-Tit-For-Tat, the traditional winners. Extort-2 was second worst, only getting a score a bit higher than Always Defect. Since evolutionary agents are not included in this tournament, extortionate strategies cannot take advantage of them. As for pairwise competition, Always Defect wins most matches, followed by Extort-2. ZDGTFT-2, Tit-For-Tat and Generous-Tit-For-Tat cannot win a single match because of their generosity.

    The tournament results can be reproduced by Vincent's python library \cite{knight2016open}.In this paper, we are able to explain the result of pairwise competitions and introduce an infinite number of strategies to win most head-to-head matches.

\section{One Memory Strategies}
     One memory strategies are strategies which base its response entirely on the outcome of the previous round. It consists of an initial state $p_0$(the probability to cooperate in the initial iteration) and a vector $\textbf{p}=\{p_1,p_2,p_3,p_4\}=\{p_{cc},p_{cd},p_{dc},p_{dd}\}$ where $p_z$ is the probability of playing c when the outcome $z$ occurred in the previous round.

     Press and Dyson \cite{press2012iterated} proved that shortest-memory player sets the rules of the game. When the game is indefinitely repeated, for any strategy of the longer-memory player Y, X's score is exactly the same as if Y had played a certain shorter memory strategy, disregarding any history in excess of that shared with X. This conclusion enables us to focus on one memory strategies.

     The probability distribution \textbf{v} on the set of outcomes is a non-negative vector with unit sum, indexed by four states, $\textbf{v}=\{v_{cc},v_{cd},v_{dc},v_{dd}\}=\{v_1,v_2,v_3,v_4\}$ and $v_1+v_2+v_3+v_4=1$. The probability of r-th iteration is noted by \textbf{v}$^r$.
    If X uses initial probability $p_0$ and strategy $\textbf{p}=\{p_1,p_2,p_3,p_4\}$, Y uses initial probability $q_0$ and strategy $\textbf{q}=\{q_1,q_2,q_3,q_4\}$, then the probability distribution of the first iteration is
    $\textbf{v} ^1=(p_0q_0,p_0(1-q_0),(1-p_0)q_0,(1-p_0)(1-q_0))$ and the successive outcomes follow a markov chain with transition matrix given by:
        \begin{equation}\label{MarkovMatrix}
         \mathbf{M} =
        \begin{pmatrix}
          p_1q_1 & p_1(1-q_1) & (1-p_1)q_1 & (1-p_1)(1-q_1) \\
          p_2q_3 & p_2(1-q_3) & (1-p_2)q_3 & (1-p_2)(1-q_3) \\
          p_3q_2 & p_3(1-q_2) & (1-p_3)q_2 & (1-p_3)(1-q_2) \\
          p_4q_4 & p_4(1-q_4) & (1-p_4)q_4 & (1-p_4)(1-q_4)
        \end{pmatrix}\nonumber
        \end{equation}

    Each entry of $\mathbf{M}$ represents the probability of transition between different states, which satisfies
    $$\mathbf{M}\mathbf{v}^r=\mathbf{v}^{r+1}$$

    In accordance with \cite{akin2013good}, we will call M convergent when there is a unique stationary distribution vector for M. Although the sequence of $\textbf{v}^i(i=1,2,...)$ may circulate among several states and thus not converge, the sequence of the Cesaro averages $\{\frac{1}{n}\sum_{i=1}^{n}\mathbf{v}^i \}$ of the outcome distributions always converges to some stationary distribution $\mathbf{v}$. That is,
    \begin{equation}\label{CesaroAverage}
        \lim_{n\to \infty}\frac{1}{n}\sum_{k=1}^{n}\mathbf{v}^k = \mathbf{v}
    \end{equation}
    In this paper, we refer to stationary distribution as Cesaro average defined in equation (\ref{CesaroAverage}).

    The following part briefly introduces Press and Dyson's conclusions of zero determinant strategies.
    The dot product of an arbitrary four-vector $\mathbf{f}=\{f_1,f_2,f_3,f_4\}$ with the stationary vector $\mathbf{v}$ of markov matrix can be written as below.
    \begin{align}
    \mathbf{v}\cdot\mathbf{f}   &\equiv D(\mathbf{p},\mathbf{q},\mathbf{f}) \nonumber \\
                                &=
                                \begin{vmatrix}
                                    p_1q_1 - 1    & p_1-1          &  q_1-1        & f_1 \\
                                    p_2q_3        & p_2-1          &  q_3          & f_2 \\
                                    p_3q_2        & p_3            &  q_2-1        & f_3 \\
                                    p_4q_4        & p_4            &  q_4          & f_4 \\
                                \end{vmatrix}  \nonumber
    \end{align}

    Following this we can calculate the average payoff of $s_X$ and $s_Y$ by $\mathbf{S_X}=(R,S,T,P)$ and $\mathbf{S_Y}=(R,T,S,P)$.

    \begin{align}\label{Normalized}
        s_{X} = \frac{\mathbf{v}\cdot \mathbf{S_X}}{\mathbf{v}\cdot \mathbf{1}} = \frac{D(\mathbf{p},\mathbf{q},\mathbf{S_X})}{D(\mathbf{p},\mathbf{q},\mathbf{1})} \nonumber  \\
        s_{Y} = \frac{\mathbf{v}\cdot \mathbf{S_Y}}{\mathbf{v}\cdot \mathbf{1}} = \frac{D(\mathbf{p},\mathbf{q},\mathbf{S_Y})}{D(\mathbf{p},\mathbf{q},\mathbf{1})}
    \end{align}

    Because the scores in Eq. (\ref{Normalized}) depend linearly on their corresponding  payoff matrices $\mathbf{S}$, the same is true for any linear combination of scores, giving
    \begin{equation}\label{linearCombination3}
        \alpha s_{X}+\beta s_{Y} + \gamma = \frac{D(\mathbf{p},\mathbf{q}, \alpha \mathbf{S_X} +\beta \mathbf{S_Y} + \gamma \mathbf{1}) }{D(\mathbf{p},\mathbf{q},\mathbf{1})}
    \end{equation}
    If X chooses a strategy $\tilde{\mathbf{p}}$  or Y chooses a strategy $\tilde{\mathbf{q}}$ that satisfies
    $ \tilde{\mathbf{p}}= \alpha \mathbf{S_X} +\beta \mathbf{S_Y} + \gamma \mathbf{1}$ or $  \tilde{\mathbf{q}}= \alpha \mathbf{S_X} +\beta \mathbf{S_Y} + \gamma \mathbf{1}$, where $\mathbf{p},\mathbf{q}\in [0,1]$,
    then the numerator in (\ref{linearCombination3}) vanishes to zero. The zero determinant of $D(\mathbf{p},\mathbf{q}, \alpha \mathbf{S_X} +\beta \mathbf{S_Y} + \gamma \mathbf{1})$ guarantees a linear relationship between the two scores,
    \begin{equation}\label{conclusion}
        \alpha s_{X}+\beta s_{Y} + \gamma = 0 \nonumber
    \end{equation}

    \textbf{Extortionate Strategy.} Let $ \tilde{\mathbf{p}} =  \phi[(\mathbf{S}_X - P \mathbf{1}) - \chi(\mathbf{S}_Y-P\mathbf{1})] $, where $\chi \geq 1$ is the extortion factor. We have
    \begin{align}\label{ExtortCondition}
      p_1 = 1-\phi(\chi-1)\frac{R-P}{P-S},& \quad p_2 = 1-\phi(1+\chi\frac{T-P}{P-S}) \nonumber \\
      p_3 = \phi(\chi+\frac{T-P}{P-S}),&    \quad p_4 = 0
    \end{align}
    Then the extortionate relationship is forced,
    $$ s_X - P = \chi(s_Y-P) $$
    $\mathbf{p}=(11/13,1/2,7/26,0)$ is concrete example of extortionate strategy with the extortion factor $\chi=3$.

    Since extortionate strategies can force an linear relationship between both players' payoffs, the co-player Y's reaction leads to disparate interesting results. If Y is an evolutionary player who adjusts his strategy \textbf{q} according to some optimization scheme designed to maximize his score $s_Y$, X can extort Y and get a higher payoff. When Y is a sentient player who knows that X is using extortionate strategy, the game results in an ultimatum game\cite{thaler1988anomalies} where Y can choose either to cooperate(then Y is extorted) or to Defect(both players get a score of P).

    Following Press and Dyson's work, we solve $ \tilde{\mathbf{p}}= \alpha \mathbf{S_X} +\beta \mathbf{S_Y} + \gamma \mathbf{1}$ in equation (\ref{linearCombination3}) and elicit the necessary condition of zero determinant strategy.
    \begin{myCorollary}
      When R=3, S=0,T=5,P=1, a zero determinant strategy necessarily satisfies
        $$    3* p_1  -  2*p_2   -  2*p_3 + p_4 - 1 = 0 $$
    \end{myCorollary}

\section{Invincible Strategies}
    \subsection{Stationary Distribution}
    The stationary vector $\mathbf{v}=\{v_1,v_2,v_3,v_4\}$ satisfies
        $$  \mathbf{M}\mathbf{v} = \mathbf{v} $$
        $$ v_1+v_2+v_3+v_4 = 1$$
    After abbreviation,
    $$
        \begin{bmatrix}
          p_1-1 & p_2-1 & p_3  & p_4\\
          q_1-1 & q_3   &q_2-1 &q_4 \\
          p_1q_1-1 & p_2q_3 & p_3q_2 & p_4q_4 \\
          1 & 1 & 1 & 1
        \end{bmatrix}
        \mathbf{v} =
        \begin{bmatrix}
          0 \\
          0 \\
          0 \\
          1
        \end{bmatrix}
    $$
   Let
    \begin{equation}\label{D}
    D =
    \begin{vmatrix}
        p_1-1 & p_2-1 & p_3  & p_4\\
        q_1-1 & q_3   &q_2-1 &q_4 \\
        p_1q_1-1 & p_2q_3 & p_3q_2 & p_4q_4 \\
        1 & 1 & 1 & 1
    \end{vmatrix}
    \end{equation}

    \begin{equation}\label{D2}
    D_2 =
    \begin{vmatrix}
        p_1-1 & 0 & p_3  & p_4\\
        q_1-1 & 0   &q_2-1 &q_4 \\
        p_1q_1-1 & 0 & p_3q_2 & p_4q_4 \\
        1 & 1 & 1 & 1
    \end{vmatrix}
    \end{equation}

    \begin{equation}\label{D3}
    D_3 =
    \begin{vmatrix}
        p_1-1 & p_2-1 & 0  & p_4\\
        q_1-1 & q_3   & 0 &q_4 \\
        p_1q_1-1 & p_2q_3 & 0 & p_4q_4 \\
        1 & 1 & 1 & 1
    \end{vmatrix}
    \end{equation}

    If $D=0$, the stationary distribution $\mathbf{v}$ is not unique, which depends on the initial distribution $\mathbf{v}^1$. For example, when both players take strategy $Repeat$, that is, $\mathbf{p}=\mathbf{q}=(1,1,0,0)$, then $D=0$ and the stationary vector $\mathbf{v}$ is exactly the same as initial distribution $\mathbf{v}^1$.

    If $D \neq 0$, there is a unique stationary distribution $\mathbf{v}$ that is independent of initial distribution $\mathbf{v}^1$. Such stationary distributions can be calculated according to Cramer's rule. We give the value of $v_2$ and $v_3$, which will be used in this paper.
    \begin{equation}\label{v2v3}
        v_2 = \frac{D_2}{D}, v_3 = \frac{D_3}{D}
    \end{equation}

    \begin{myTheorem}
        Assume $p_1,...,p_4,q_1,...,q_4\in[0,1]$, $D$ is defined in equation (\ref{D}),
        $$D \leq 0$$
    \end{myTheorem}
    \begin{myProof}
        $\forall z\in \{p_1,...,p_4,q_1,...,q_4\}$,
        \begin{equation}\label{partialpartialDis0}
          \frac{\partial^2 D}{\partial^2 z} = 0
        \end{equation}
        Equation (\ref{partialpartialDis0}) implies that, when all variables in  $\{p_1,...,p_4,q_1,...,q_4\}\backslash z$ are fixed, $$\frac{\partial D}{\partial z} = C$$
        where $C$ is a constant independent of $z$. In this case, $D$ is monotonous to $z$ when other variables are fixed.

        Therefore we can get all extreme values of $D$ by letting
        $$z=0\quad or \quad z=1,\quad \forall z\in \{p_1,...,p_4,q_1,...,q_4\}$$

        Since all of $2^8=256$ extreme values are less or equal to zero, we can conclude that $D\leq 0$. \qed
    \end{myProof}

    \subsection{Invincible Strategies}
    \begin{myDefinition}[Invincible Strategy]
        A one memory strategy \textbf{p} is invincible if against any
other one memory strategy \textbf{q}, $s_X\geq s_Y$, provided $s_X$ and $s_Y$
exist according to (\ref{Normalized}).
    \end{myDefinition}
Notice that according to our definition, a one memory strategy \textbf{p}
is invincible
if it never lose against other one memory strategies. However, according to
a result by Press and Dyson \shortcite{press2012iterated}, this also
mean that \textbf{p} never lose against any strategies as having longer memory
will not help.

\begin{myTheorem}
Assume $s_X$ and $s_Y$ exist according to (\ref{Normalized}), and that $\mathbf{v}=\{v_1,v_2,v_3,v_4\}$ is the unique stationary vector, then
    $$s_X \geq s_Y \Longleftrightarrow v_2 \leq v_3 $$
\end{myTheorem}
\begin{myProof}
   \begin{align}\label{Proof2}
      & s_X \geq s_Y  \nonumber \\
     \Longleftrightarrow \quad & \mathbf{v}\cdot \mathbf{S_X} \geq \mathbf{v}\cdot \mathbf{S_Y} \nonumber \\
     \Longleftrightarrow \quad & S*v_2 + T*v_3 \geq T*v_2 + S*v_3  \nonumber \\
     \Longleftrightarrow \quad & (T-S)*(v_3-v_2) \geq 0  \nonumber \\
     \Longleftrightarrow \quad & v_2 \leq v_3   \nonumber
   \end{align}\qed
\end{myProof}

\begin{myTheorem}
	Invincible strategy $\mathbf{p}=(p_1,p_2,p_3,p_4)$ must satisfy $p_4=0$.
\end{myTheorem}
\begin{myProof}
	Suppose $p_4>0$, when $\mathbf{p}$ plays against Always Defect, from \textbf{p}'s perspective, the terminal states of this game should be $\{dd,cd\}$ or $\{cd\}$. In both cases, strategy \textbf{p} is defeated by strategy Always Defect. \qed
\end{myProof}

Our main result is as follows:
\begin{myTheorem}
    $\textbf{p}=(p_1,...,p_4)$ is invincible iff
\begin{equation}\label{p2p3lt1}
    p_2+p_3 \leq 1
\end{equation}
   $$ p_4 = 0 \nonumber $$
\end{myTheorem}

\begin{myProof}
    Suppose player X takes strategy \textbf{p} while player Y takes strategy \textbf{q}. In this proof we assume $D \neq 0$, the edge cases $D=0$ will be discussed case by case later.\\
\textbf{Necessity.}
    Since $\mathbf{p}$ is invincible,
    $$ \forall \textbf{q}, s_X \geq s_Y $$

    According to Theorem 3, \textbf{p} is invincible, thus
    \begin{equation}\label{proofp4is0}
        p_4 = 0
    \end{equation}

    According to Theorem 2,
    $$ \forall \textbf{q}, v_2 \leq v_3 $$

    Because we suppose $D \neq 0$, according to (\ref{v2v3}),
	\begin{equation}
        \forall \textbf{q}, D_2 \geq D_3
        \label{D2geqD3}
    \end{equation}

    Since $p_4=0$, according to equation (\ref{D2})(\ref{D3}),
    \begin{align}
      D_2 &= p_3q_2q_4 - p_3q_4 + p_1p_3q_1q_4 - p_1p_3q_2q_4 \nonumber \\
      D_3 &= p_2q_4 - q_4 + p_1q_1q_4 - p_2q_3q_4 - p_1p_2q_1q_4 + p_1p_2q_3q_4 \nonumber
    \end{align}
    Let $\mathbf{q}=(0,0,0,1)$,
    \begin{equation}
      D_2 = - p_3, D_3 = p_2 - 1
    \end{equation}

    According to (\ref{D2geqD3}),$D_2 \geq D_3$, we have
    $$p_2 + p_3 \leq 1 $$

    With the addition to (\ref{proofp4is0}), we prove the necessity.\\ \\
    \textbf{Sufficiency.}
    \begin{align}\label{AfterMerge}
    Let \quad \mathcal{L} = &(1-p_2-p_3)(1-p_1q_1)   \nonumber \\
      &+ (1 - p_1)p_3q_2 + (1-p_1)p_2q_3    \nonumber
    \end{align}
    Given that $p_1,...,p_4,q_1,...,q_3 \in [0,1]$, and that we assume $p_2 + p_3 \leq 1$ in equation (\ref{p2p3lt1}), we have
    $$\forall q_1,...,q_3 \in [0,1],\quad \mathcal{L} \geq 0$$
    Because we assume $p_4=0$ and $0\leq q_4 \leq 1$, notice that
    $$ q_4 \cdot \mathcal{L} = D_2 -D_3 $$
    We have
    $$ \forall \mathbf{q},\quad D_2 \geq D_3 $$
    According to Theorem 1, $D \leq 0$, and we assume $D\neq 0$, we have
    $$\forall \mathbf{q},\frac{D_2}{D} \leq \frac{D_3}{D}$$
    According to Equation (\ref{v2v3}), it's the same as,
    $$\forall \mathbf{q}, v_2 \leq v_3 $$
    According to Theorem 2,
    $$\forall \mathbf{q}, s_X \geq s_Y $$
    Therefore, strategy \textbf{p} is invincible.
          \qed
    \end{myProof}

     \begin{myTheorem}
        All extortionate strategies are invincible strategies.
    \end{myTheorem}
    \begin{myProof}
        According to equation set (\ref{ExtortCondition}), extortionate strategies satisfy $p_4=0$.
        \begin{equation}\label{ExtortIsInvincible}
          p_2+p_3 = 1 + \phi(1-\chi)(\frac{T+S-2P}{P-S})
        \end{equation}
        Recall the setting of IPD, $T+S > 2P$ and $P>S$. In extortionate strategies, $\chi \geq 1, \phi > 0$. Then we have $p_2+p_3 < 1$.

        Therefore, all extortionate strategies are invincible. \qed
    \end{myProof}

    Invincible strategies account for a large proportion of all strategies. Half of firm strategies($p_4=0$) are invincible since the hyper plane $p_2+p_3 \leq 1$ bisects the 3D cube $(p_1,p_2,p_3)\in [0,1]$ when $p_4=0$. All extortionate strategies are invincible so that agents cannot get rid of being extorted and retaliate the extortioner with a win. Tit-for-Tat$(1,0,1,0)$, with $p_2+p_3=1$, is a special invincible strategy which equalizes the payoff of both players. By playing Defect in the first iteration, Tit-for-Tat can avoid being defeated in the beginning. Although Tit-for-Tat is invincible, it cannot defeat other strategies, either. Always Defect is an extreme invincible strategy but we now know that invincible strategies are not restricted to Always Defect, which gives us more space to explore high-score and invincible strategies. Such strategies cannot be targeted by specially designed natural enemies.

    \subsection{Edge Cases}
        When $D=0$, there are more than one terminal sets, and the stationary distribution depends on the initial one, which occurs when some of $\{p_1,...,._4,q_1,...,q_4\}$ equal to 0 or 1. As iterated prisoner's dilemma is usually deemed as playing for infinite rounds, or at least an unknown large number of rounds, the final distribution shouldn't depend much on the initial one \cite{sigmund2010calculus}.

        However, to be rigorous, we enumerate all edge cases when $D=0$ and analyse them case by case.

        According to Theorem 4, we only analyze the cases where $p_2 + p_3 \leq 1$ and $p_4=0$. The following discussion is from \textbf{p}'s perspective, i.e. $\{cd\}$ refers to the state where \textbf{p} plays $c$ and \textbf{q} plays $d$.\\
        \textbf{Case 1.} \textbf{p}=(0,0,0,0). Obviously, no matter what \textbf{q} is, {\it Always Defect} is invincible. \\
        \textbf{Case 2.} \textbf{p}=(0,0,1,0). Because $\{cd\}$ can only appear exactly after $\{dc\}$ except for this first iteration, this strategy is invincible if the game is played for infinite rounds. \\
        \textbf{Case 3.} \textbf{p}=(0,1,0,0). This strategy is \textbf{NOT} invincible since $\{cd\}$ can be a stationary distribution.  \\
        \textbf{Case 4.} \textbf{p}=(1,0,0,0). {\it Trigger Strategy} is invincible since $\{cd\}$ can only appear at most once in game history and we assume the game is played for infinite rounds.  \\
        \textbf{Case 5.} \textbf{p}=(1,0,1,0). {\it Tit for Tat} is invincible since $\{cd\}$ can only appear exactly after $\{dc\}$, except for the first time after $CC$ or the first iteration, which doesn't matter when the game is played for infinite rounds.  \\
        \textbf{Case 6.} \textbf{p}=(1,1,0,0). Obvious, {\it Repeat} is \textbf{NOT} invincible if it plays $C$ in the first iteration.\\
        \textbf{Case 7.} $\mathbf{q}=(1,1,0,0)$. {\it Repeat} is the only situation where \textbf{q} can unilaterally set $D=0$ when $\forall z \in \{p_1,p_2,p_3,p_4\}$, $z\neq 0$ and  $z\neq 1$.

        All edge cases can be excluded by playing {\it Defect} in the first iteration. In this situation, \textbf{p} becomes {\it Always Defect } in case 1,3,4 and 6, while in case 2 and 4, $\{cd\}$ can only appear exactly after $\{dc\}$. As for case 7, all outcomes become $\{dd\}$.

        We can also assume $p_2 \neq 1$ and play {\it Cooperate} in the first iteration , which excludes case 3 and case 6. In case 1,2,4,5, in the infinite sequence of game history, there is one more \{cd\} than \{dc\}. In terms of stationary distribution and average payoff, losing the shot game once doesn't make any difference. As for case 7, suppose \textbf{q} is {\it Always Defect}, \{dd\} is the only absorbing state while \{cd\} is only a transient state, thus strategy \textbf{p} is still invincible.

    \subsection{Stochastic Calculation}
\begin{figure}[]
\subfigure[\textbf{p}=(0.5,0.2,0.7,0)]{
\centering
\includegraphics[width=0.2\textwidth]{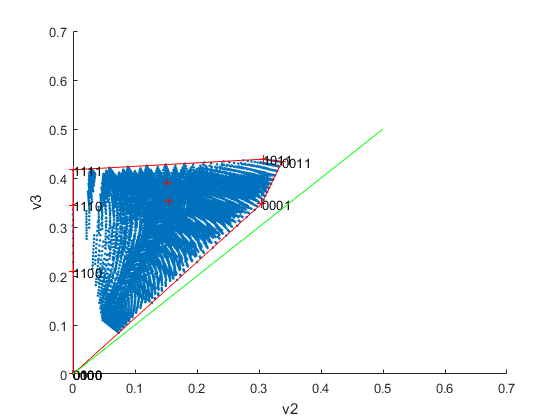}
\label{StochasticCalculationa}}
\hspace{0cm}
\subfigure[\textbf{p}=(0.5,0.7,0.2,0)]{
\centering
\includegraphics[width=0.2\textwidth]{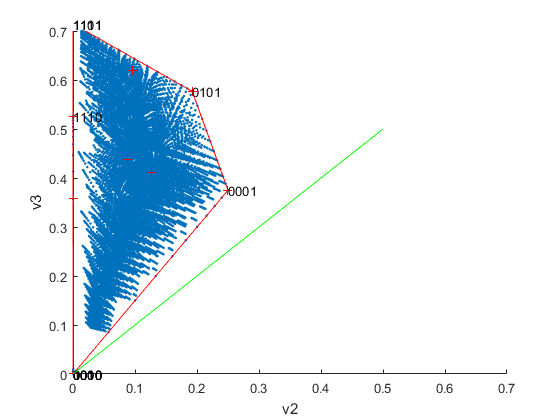}
\label{StochasticCalculationb}}
\subfigure[\textbf{p}=(0.5,0.7,0.8,0)]{
\centering
\includegraphics[width=0.2\textwidth]{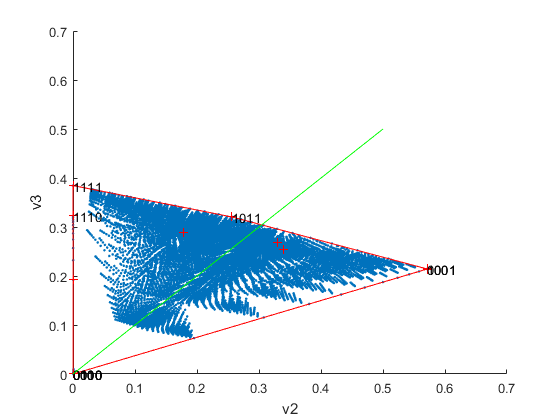}
\label{StochasticCalculationc}}
\hspace{0.85cm}
\subfigure[\textbf{p}=(0.5,0.7,0.2,0.4)]{
\centering
\includegraphics[width=0.2\textwidth]{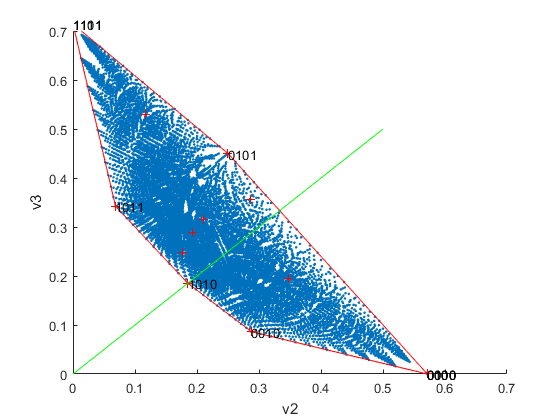}
\label{StochasticCalculationd}}
\caption{Stochastic Calculation}
\label{StochasticCalculation}
\end{figure}

     Figure \ref{StochasticCalculation} displays the result of stochastic calculation. Given a strategy $\mathbf{p}$, enumerate the strategy of $\mathbf{q}$ and display all of the stationary distributions in terms of $v_2$ and $v_3$. Because all distributions form a compact convex hull\footnote{Due to the limited space, we don't give a formal proof of this property.}, we can make discrete samples to get the shape. According to Akin's Lemma\cite{akin2013good} and the property $v_1+v_2+v_3+v_4=1$, we can represent $v_1,v_4$ by $v_2,v_3$ and visualize the final results in two dimensional plane when strategy \textbf{p} is fixed.
     \begin{myLemma}[Akin's Lemma]
        Assume that X uses the strategy \textbf{p} with X Press-Dyson vector $\mathbf{\tilde{p}}=\{p_1-1,p_2-1,p_3,p_4\}$. If the opponent Y uses a strategy pattern that yields a sequence of distributions $\{\mathbf{v}^n\}$, then
    \begin{equation}
        \lim_{n\to \infty}\frac{1}{n} \sum_{k=1}^{n}<\mathbf{v}^k \cdot \tilde{\mathbf{p}}>  = 0 \nonumber
    \end{equation}
    \begin{equation}\label{AkinsLemma}
        and\quad so \quad <\mathbf{v} \cdot \tilde{\mathbf{p}}> = v_1 \tilde{p_1} + v_2 \tilde{p_2}+ v_3 \tilde{p_3}+ v_4 \tilde{p_4} = 0. \nonumber
    \end{equation}
    \end{myLemma}

     Figure \ref{StochasticCalculationa} and figure \ref{StochasticCalculationb} are positive examples. As is shown, all results are above the line of $v_2=v_3$, which means $v_3 \geq v_2$ in all stationary distributions. The result of figure \ref{StochasticCalculationb} is counterintuitive because $p_{cd}=0.7$, which means that even if it was defected in the previous round, its probability of cooperation is as high as 0.7. Although it seems to be easily defeated, it is still invincible because it satisfies Theorem 4. Figure \ref{StochasticCalculationc} is a counter example where $p_2+p_3 > 1$. For this kind of strategy, Always Defect can merely reach a tie of Punishment, but strategies like $(0,0,0,1)$ can defeat it and thus get a higher payoff. Figure \ref{StochasticCalculationd} is a counter example where $p_4 > 0$, which can be defeated by Always Defect.

\section{Experiment}
    We have modeled IPD as a Markov Process and calculated stationary distribution. In this section, we use agent based modeling(ABM) to take experiments. We research in the following questions,
    \begin{itemize}
      \item Does the result of ABM accord with stochastic calculation?
      \item How many iterations does it approximately take to reach the stationary distribution?
      \item Can invincible strategies defeat other strategies besides one memory strategies?
      \item What's the evolutionary behavior of invincible strategies?
    \end{itemize}

\subsection{Pairwise Competition}
    \begin{figure}
    \subfigure[\textbf{p}=(0.5,0.2,0.7,0)]{
    \centering
    \includegraphics[width=0.22\textwidth]{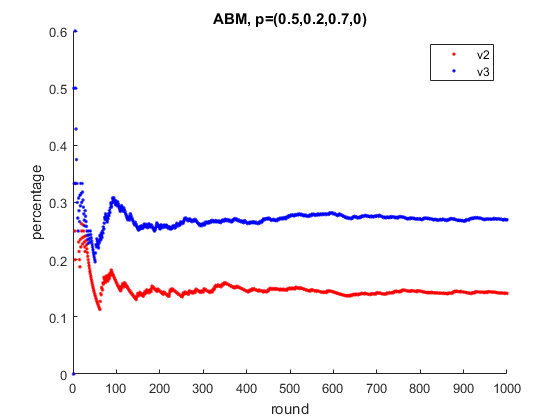}
    \label{ABM1a}}
    \subfigure[\textbf{p}=(0.5,0.7,0.2,0)]{
    \centering
    \includegraphics[width=0.22\textwidth]{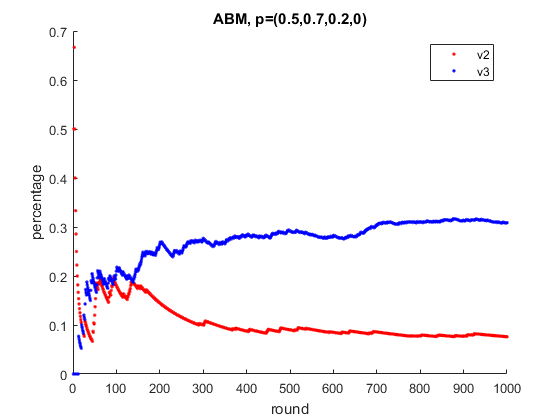}
    \label{ABM1b}}
    \caption{ABM Result}
    \label{ABM}
    \end{figure}

   \begin{table}
      \centering
      \begin{tabular}{|c|c|c|}
        \hline
          & \textbf{Stochastic Calculation} & \textbf{ABM(1000 rounds)} \\
        \hline
         $\mathbf{p_1}$ & (0.142,0.150,0.272,0.437) & (0.152,0.141,0.270,0.437) \\
        \hline
         $\mathbf{p_2}$ & (0.080,0.077,0.314,0.529) & (0.076,0.076,0.309,0.539) \\
        \hline
      \end{tabular}
      \caption{Stationary Distributions} \label{stationaryTable}
    \end{table}

    We use Java to model the behavior of agents in IPD. All agents take one memory strategies and outcomes of every round are recorded. Two invincible strategies $\mathbf{p_1}=\{0.5,0.2,0.7,0\}$ and $\mathbf{p_2}=\{0.5,0.7,0.2,0\}$ play against a specific normal strategy $\mathbf{q}=\{0.4,0.5,0.6,0.3\}$ respectively. The game is played for 1000 rounds. The percentage of each state is recorded after each round and the percentage of $v_2,v_3$ is shown in Figure \ref{ABM}. The distribution gradually becomes stable after approximately 500 and 750 rounds respectively.
    Table \ref{stationaryTable} compares the result of stochastic calculation(see section Stationary Distribution) and agent-based modeling. After 1000 iterations, the result of ABM has almost converged to the result of stochastic calculation.

    Look back at the Stewart's tournament.
    The result that Extort-2 can be ranked in second place with respect to pairwise competitions is not due to its extortionate property, but to its invincible property. Extortionate strategies are only special when it is faced with evolutionary agents. To verify this, we rerun this tournament with the help of Axelrod Python library\cite{knight2016open}. We replace Extort-2 strategy with an invincible strategy (1,0.7,0.2,0),who plays $D$ in the first iteration. According to Corollary 1, (1,0.7,0.2,0) is not a zero determinant strategy, nor is extortionate strategy.

    Each pairwise game is played for 1000 iterations to get a more accurate approximation of stationary distribution. The result of head-to-head matches is displayed in table \ref{TournamentWithInvin}. Compared to Extort-2 in Stewart's tournament, this invincible strategy has higher average payoff (1.60 vs. 1.58) and wins more head-to-head matches(16 vs. 14). Therefore, we can conclude that extortionate strategies have no advantage in Axelrod's tournament other than it is an invincible strategy.

    Stewart's tournament contains not only one memory strategies. Tit-For-Two-Tats is a two memory strategy, Prober is a stochastic strategy and GoByMajority even examines the entire history. To some extent, this tournament result verifies our assumption that shortest memory players set the rule of the game, which enables us to focus on one memory strategies.

    \begin{table}
      \centering
      \begin{tabular}{|c|c|c|}
      \hline
            \textbf{Strategy} & \textbf{Score} & \textbf{Invincible vs.} \\
        \hline
            ZD-GTFT-2 & 2.79 & win \\
        \hline
            GTFT & 2.59 & win \\
        \hline
            Tit For Tat(C) &  2.54  & win \\
        \hline
            Tit For 2 Tats &  2.41  & win \\
        \hline
            Hard Prober &  2.39  & win \\
        \hline
            Win-Stay Lose-Shift & 2.32  & win \\
        \hline
            Hard Tit For 2 Tats  & 2.30  & win \\
        \hline
            Random &  2.11  & win \\
        \hline
            Cooperator  & 2.07  & win \\
        \hline
            Prober  &  2.01  & win \\
        \hline
            Grudger & 2.00  & win \\
        \hline
            Hard Go By Majority  & 2.00  & tie \\
        \hline
            Hard Tit For Tat  & 1.99  & win \\
        \hline
            Calculator  & 1.90  & win \\
        \hline
            Prober  & 1.86  & win \\
        \hline
            Prober  & 1.77   & win \\
        \hline
            Joss    &  1.69   & win \\
        \hline
            Invincible Strategy &  1.60   & - \\
        \hline
            Defector  & 1.57   & tie         \\
        \hline
      \end{tabular}
      \caption{Tournament with Invincible Strategies} \label{TournamentWithInvin}
    \end{table}

\subsection{Evolutionary Behaviour}
\begin{figure}[thbp]
        \subfigure[WSLS, Defector]{
        \centering
        \includegraphics[width=0.22\textwidth]{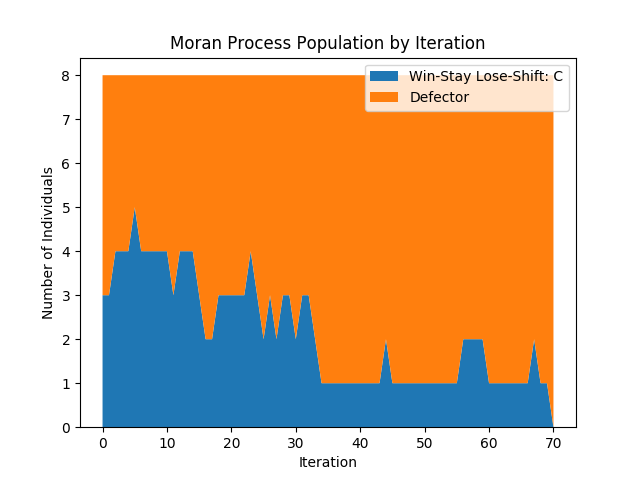}
        \label{WSLSDefector}}
        \subfigure[WSLS, Defector, Invincible]{
        \centering
        \includegraphics[width=0.22\textwidth]{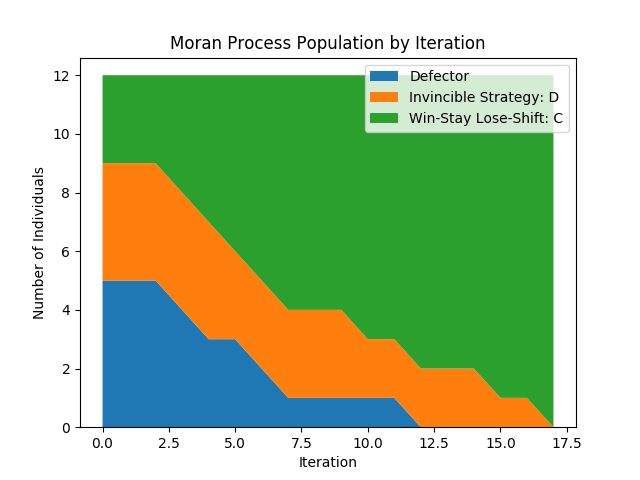}
        \label{WSLSDefectorInvin}}
        \subfigure[One Population]{
        \centering
        \includegraphics[width=0.22\textwidth]{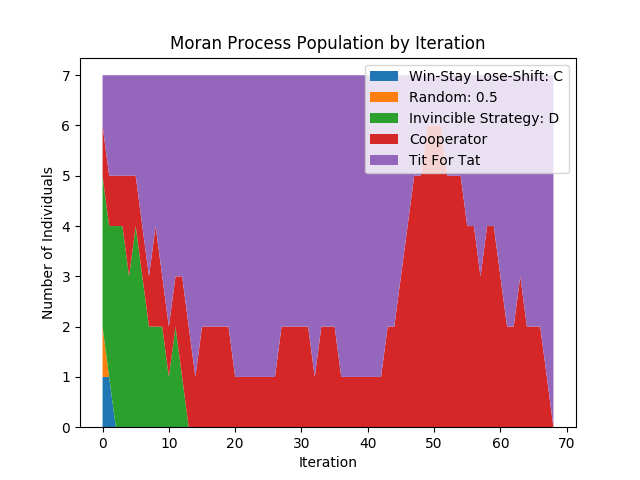}
        \label{OnePopulation}}
        \subfigure[Two Allies]{
        \centering
        \includegraphics[width=0.22\textwidth]{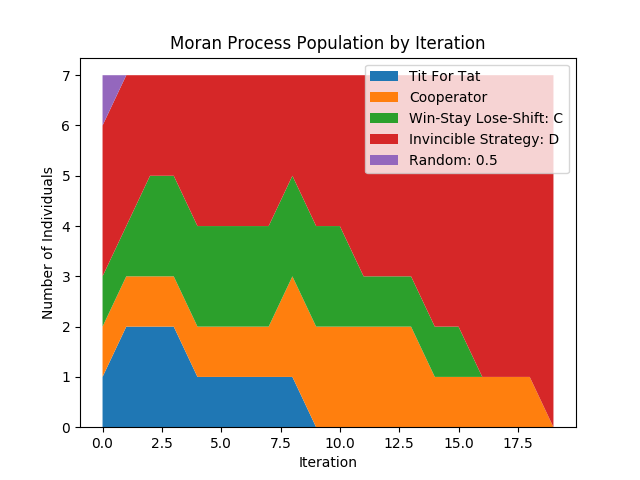}
        \label{TwoAllies}}
        \caption{Evolutionary Results}
        \label{EvolutionaryResults}
    \end{figure}

    Hilbe {\it et al.} \cite{hilbe2013evolution} analyzed the evolutionary performance of extortionate strategies. They concluded that extortionate strategies can act as catalysts for the evolution of cooperation but that they are not the stable outcome of natural selection. We rerun their experiment and replace extortionate strategy with our invincible strategy (0.9,0.7,0.2,0), and the result turns out to be similar. In Axelrod’s original work\cite{hamilton1981evolution} an ecological approach based on the payoff matrix of the tournament was used to study the evolutionary stability of each strategy. We take Moran Process\cite{lieberman2005evolutionary} in our experiment, because it is much more widely used in the literature. In Figure \ref{WSLSDefector}, win stay lose shift(WSLS) strategy is dominated by Defector and finally dies out. After adding some invincible agents, WSLS dominates this population(Figure \ref{WSLSDefectorInvin}).

    To explain this experimental result, we define {\it Semi-Cooperative Invincible Strategy}.
    \begin{myDefinition}[Semi-Cooperative Invincible]
        A strategy $\textbf{p}=(p_1,...,p_4)$ is semi-cooperative invincible if it plays {\it Cooperate} in the first iteration and satisfies,
        $$ 0.5 < p_1 <1  $$
        $$ p_2+p_3 \leq 1 $$
        $$ p_4 = 0 \nonumber $$
    \end{myDefinition}
    Such semi-cooperative invincible strategies can act as catalyst of cooperation(Figure \ref{WSLSDefector},\ref{WSLSDefectorInvin}) because it has the following properties when the game is played for a finite number of rounds. (1) When such strategies play against defective strategies such as {\it Always Defect}, the distribution will quickly fall into {\it DD}, thus both players receive a reward of {\it P}. (2)When such strategies play against cooperative strategies, such as {\it WSLS} and {\it Cooperate}, the distribution will hold at $CC$ for a number of rounds, resulting a payoff of $R$. (3)When such strategies play among themselves, it's better than defective but worse than cooperative so that the average payoff is between (1) and (2). As a result, when the agent with lowest payoff is eliminated, {\it Always Defect} has the least average payoff, followed by semi-cooperative invincible strategy, and finally cooperative strategy. Such property make semi-cooperative invincible strategy seem like a catalyst of cooperation. 

    Following Hilbe's discussion of extortionate strategy between two populations\cite{hilbe2013evolution}\cite{hilbe2014cooperation}, we take similar experiment to simulate the performance of invincible strategies when they form an alliance. In a single population invincible strategies need to compete with their own kind, which decreases their payoff. However, when invincible strategies evolve in one of two separate populations, they will show their battle effectiveness. We still take Moran Process and build a bipartite graph of players\cite{shakarian2013novel}, of which one party contains invincible players while the other party contains other strategies. When all players in one population, cooperative strategies such as Cooperator and Tit for Tat dominates this society(Figure \ref{OnePopulation}). After invincible strategies forms an ally and play against the other population, they dominate this ecosystem(Figure \ref{TwoAllies}).

\section{Conclusion and Future Work}
    Inspired by the fact that no strategies can defeat extortionate strategy, we discovered invincible strategies which will not lose head-to-head matches if the game is played for a large enough number of iterations. We give a concise characterization of such strategies and verifies it by experiments.

    Replacing extortionate strategy with some other invincible strategies, we conduct tournaments and evolutionary experiments which aim to analyze the similarity between extortionate strategy and invincible strategy. Our results show that the properties of extortionate strategies discovered by recent works, such as winning head-to-head matches and acting as catalyst for cooperation, are actually the properties of some invincible strategies. Extortionate strategies are only special when it plays with an evolutionary player so that it can direct the co-player's evolutionary direction.

    In addition to reproducing experiments in related works, we also give mathematical insights of such phenomena, which explains why extortionate strategies can win head-to-head matches by invincible condition(Theorem 4), illustrates why it can be catalyst of cooperation in one population(Definition 2) and dominant another population.

    Although winning pairwise competition is not everything\cite{adami2013evolutionary}, keeping the invincible condition in mind, sentient agents are able to evaluate the vulnerability of the co-player. Given that invincible strategies are not restricted to Always Defect, it's now possible to explore high-payoff and invincible strategies. The invincible property guarantees the lower bound of agents' payoff, which improves environmental adaptability of agents.

    In future works, we will analyze why generous zero determinant strategy can achieve the maximum score in Stewart's tournament. Current research of scores requires a specific distribution of co-players, which is not stable when experiment configuration varies. We will continue our research of one memory strategies and focus on the average payoff of pairwise competitions. We hope that our research can take more insights into iterated prisoner's dilemma.

\section{ Acknowledgments}
Acknowledgement.

\bibliography{AlgrithmicGameTheory}
\bibliographystyle{aaai}

\end{document}